# The Dark Universe Is Not Invisible [†]

K. Zioutas[1], V. Anastassopoulos[1], A. Argiriou[1], G. Cantatore[2],
S.A. Cetin[3], A. Gardikiotis[1,‡], D.H.H. Hoffmann[4], S. Hofmann[5], M. Karuza[6],
A. Kryemadhi[7], M. Maroudas[1,*], E.L. Matteson[8], K. Ozbozduman[9],
T. Papaevangelou[10], M. Perryman[11], Y.K. Semertzidis[12,13], I. Tsagris[1,§],
M. Tsagri[1,§], G. Tsiledakis[10], D. Utz[14] and E.L. Valachovic[15]

[1] Physics Department, University of Patras, GR 26504 Patras-Rio, Greece
[2] Physics Department, University & INFN of Trieste, 34127 Trieste, Italy
[3] Institute of Sciences, Istinye University, 34396, Istanbul, Turkey
[4] School of Science, Xi'an Jiaotong University, Xi'an 710049, China
[5] Independent Researcher, Fastlinger Strasse 17, 80999 München, Germany
[6] Physics Department, University of Rijeka, 51000 Rijeka, Croatia
[7] Physics Department, Messiah University, Mechanicsburg, PA 17055, USA
[8] Mayo Clinic, Rochester, MN 55902, USA
[9] Physics Department, Boğaziçi University, 34342 Istanbul, Turkey
[10] IRFU/CEA-Saclay, CEDEX, 91191 Gif-sur-Yvette, France
[11] School of Physics, University College Dublin, Belfield, Dublin 4, Ireland
[12] Center for Axion and Precision Physics Research, Institute for Basic Science, Daejeon 34051, Korea
[13] Department of Physics, Korea Advanced Institute of Science and Technology, Daejeon 34141, Korea
[14] Institute for Geophysics, Astrophysics and Meteorology, University of Graz, 8010 Graz, Austria
[15] Department of Epidemiology and Biostatistics School of Public Health, University at Albany, State University of New York, New York, NY 12144, USA

[*] Author to whom correspondence should be addressed: **marios.maroudas@cern.ch** .
Present addresses: [‡] University of Hamburg, 20146 Hamburg, Germany. [§] Geneva, Switzerland.

**Abstract**

Evidence of dark matter (DM) comes from long-range gravitational observations, where it is understood actually to not interact with ordinary matter. However, on a much smaller scale, a number of unexpected phenomena contradict this idea of DM, since some solar activity and the dynamic Earth atmosphere might arise from DM streams. Gravitational (self-)focusing effects by the Sun or its planets of streaming DM fit as the underlying process, e.g., for the otherwise puzzling 11-year solar cycle, the mysterious heating of the solar corona with its fast temperature inversion, etc. Observationally driven, we arrive to suggest an external impact by as yet overlooked "streaming invisible matter", which reconciles some of the investigated mysterious observations. Unexpected planetary relationships exist for the dynamic Sun and Earth atmosphere and are considered as *the* signature for streaming DM. Then, focusing of DM streams could also occur in exoplanetary systems, suggesting for the first time the carrying out of investigations by searching for the associated stellar activity as a function of the exoplanetary orbital phases. The entire observationally driven reasoning is suggestive for highly cross-disciplinary approaches that also include (puzzling) biomedical phenomena. Favoured candidates are the highly ionizing anti-quark nuggets, magnetic monopoles and dark photons.

## 1. Introduction

The detection of the constituents of dark matter (DM) is one of the central challenges in modern physics. The strongest evidence of DM comes from large scale observations, while direct and indirect searches are followed by a large number of experiments. The study of anomalous phenomena in physics has provided some surprises. For example, the observation of an unexpected atmospheric ionization (1912) resulted in the discovery of cosmic rays [1]. The discovery of dark matter (DM) by Zwicky in 1933 [2] was due to gravitational discrepancies observed in large cosmological systems. The search for the direct detection of the putative DM constituents has continued for decades, though without success. DM became synonymous with the idea of something that actually does not interact with ordinary matter, and, more specifically, that does not emit light.

In this work, we point out a number of striking observations made in our neighbourhood that contradict this picture of DM [3,4,5]. These relevant observations cover diverse mysterious phenomena, with the mostly striking ones being the multifaceted solar activity, the dynamic Earth atmosphere and diseases such as cancer [6]. A common remarkable signature in all of these studies is the observation of planetary relationships that are not expected within known physics. We recall that already as early as 1859 WOLF [7] suspected planetary influence at the origin of the otherwise mysterious 11-year solar cycle, which is present in a plethora of phenomena. Meanwhile, one is inclined to accept this as something obvious. However, it was concluded observationally [3,4,5] that planetary focusing effects by some type of low-speed invisible streaming matter could be behind some of the observations. Though, lensing effects even by the Sun are effective some 545 A.U. downstream.

It is worth mentioning here that planetary gravitational lensing effects from a planet is possible within the solar system from one solar body to another one, and most notably from the Sun itself. This is so, because the gravitational deflection depends on $1/(velocity)^2$, and streaming DM constituents with velocities of around $10^{-3}c$ or less can become strongly influenced over distances typical for the inner solar system, including the Moon–Earth distance [8,9]. In between, SOFUE [10] concluded that even the intrinsic Earth can strongly gravitationally enhance the flux of incident streams downstream on Earth's opposite surface, in a velocity range of up to about 400 km/s covering a large fraction of the expected one for DM. The work of ref. [10] expands the DM scenarios of planetary lensing effects [8,9] between solar system bodies including the Earth–Moon system. The flux enhancement by the Moon towards Earth is about $10^4$ (Manuscript in preparation, A. Kryemadhi, M. Vogelsberger, K. Zioutas).

Notably, there is no known force that can explain remote planetary effects except the widely suspected gravitational tidal forces such as those acting on Earth by the Moon. However, this tidal force is far too feeble to cause solar phenomena, with a strength weaker by a factor of about $10^{-12}$ [11]. Nevertheless, the aforementioned planetary lensing effects of streaming DM constituents result in a seemingly remote interaction within our solar system. Following this mechanism, it is natural to expect planetary dependence of various phenomena that should not occur within conventional physics of forces associated between solar system bodies. Streaming DM is required for this to happen, which, in addition, must interact with ordinary matter with a much larger cross-section compared to the limits derived so far for DM axions and WIMPs. Meanwhile, DM particles with similar properties are being discussed in the literature; we mention as possible examples the theoretically motivated anti-quark nuggets (AQNs) [12] or other DM clusters, and eventually also magnetic monopoles. Following various observations, these types of particles, including the hidden sector dark photons, remain as the favorites. Since 2017, ZHITNITSKY and collaborators have elaborated on the involvement of AQNs which might explain solar phenomena such as the mysterious solar corona heating source [13], but also anomalous high energy events and other observations in space.

Thus, the planetary DM scenario already fits several observations, suggesting independently streaming DM in our neighborhood. Interestingly, DM streams [14] or clusters [15] are also motivated cosmologically.

The typical and common feature of a number of otherwise unexpected observations is their planetary relationship [3,4,5]; their manifestation is given by periodic rates of certain observables identical with those of fixed planetary orbital, synodic or other combined planetary periodicities. Therefore, to extend this approach in other observations, long time series of some measured observables are required. After several data analyses and a number of consistency tests, two developed analysis software codes are reliable.

Notably, the planetary scenario is completely different from the models based on tidal forces, which have been attempted with very little success since the discovery of the first large flare some 155 years ago [7]. However, it is worth stressing that there is no conventional explanation for a remote planetary interaction with the Sun's or Earth's atmosphere, e.g., by gravitational tidal forces, since they are by far too weak [11]. Presently, we are making neither an assumption about the nature of the streaming invisible massive matter nor about its interaction with normal matter e.g., of the Sun or Earth. Our goal is to find without bias both the lensing and the existence of preferred direction(s). If this seminal idea holds, there will be ways to explore it further in the future, due to its apparent implications in ongoing dark matter searches.

In addition, the observation of a peaking planetary relationship excludes any remote tidal forces, since their strength changes smoothly during one orbit [11]. Thus in order to identify the origin of a signature possibly showing an 11-year rhythm, the search for a planetary relationship is essential. The driving idea behind this study is based on the gravitational focusing by the Sun and its planets of low-speed invisible (streaming) matter. Whatever its ultimate properties, it must somehow interact "strongly" with normal material such as that of the upper atmosphere or the Sun's atmosphere in order to be able to cause the observed puzzling behaviour located there. Occasionally, we refer to generic dark matter constituents as "invisible massive matter", in order to distinguish them from the widely addressed dark matter candidates such as axions or WIMPs, which cannot have any noticeable atmospheric effect. Encouragingly, recent work discusses potential constituents from the dark sector [12,13] having a large cross-section with normal matter.

## 2. Experiments—Observations

A number of terrestrial and celestial observations have been analyzed in previous work [3,4,5], which have shown that at least some solar and terrestrial observations follow planetary relationships. Most of these observations were already considered to be of unknown origin, such as the flaring Sun, the solar corona paradox or the anomalous annual stratospheric temperature excursions that take place around December - January, and, atmospheric ionization. Remarkably, the latter coincides with the annual alignment [4] around 18 December of Earth, the Sun and the galactic center, which also occasionally includes the Moon.

Notably, a planetary relationship has also been observed for melanoma incidence [16], which is a type of cancer of the skin. Interestingly, this first observation has been cross-checked independently (see ref. [17]). In spite of the first unfortunate interpretation, the Fourier analysis of the same dataset remarkably provided a clear peak at the orbital period of Mercury for a number of major cancer types. This is not surprising within the advocated invisible streaming matter scenario, given the inherent sensitivity of living matter to external influence. Moreover, recently, the analysis of a long series of daily melanoma incidence registered in Australia provided a short periodicity of 27.3 days [6]. This rhythm coincides exactly with the Luna orbital period fixed to remote stars, which implies that its origin, whatever it is ultimately found to be, must be exo-solar. The aforementioned invisible streaming matter scenario fits within this construct, since the Moon can focus DM constituents towards Earth with velocities of up to about 400 km/s [10], which covers a large portion of the DM's velocity distribution, peaking at about 250 km/s.

## 3. Results

Here we add two more solar observables [18]. The same analysis as before (**Section 2**) has been applied with daily measurements. The two solar observables are:

(a) the elemental composition of the Sun's atmosphere [19]. Surprisingly, the elemental composition in the corona and slow solar wind ($A_c$) is different than in the photosphere ($A_P$). Hence, the values $A_c/A_p$ provide the ratio of the coronal abundance ($A_c$) to the photospheric abundance ($A_p$). This composition enhancement process is known as the FIP effect. Low first ionization potential (FIP) elements that are easy to ionize in the chromosphere are preferentially enhanced by factors of 3- to 4-fold, whereas high-FIP elements that remain neutral in the chromosphere retain their photospheric abundances. The variation of coronal composition is highly correlated with solar activity as it is given by the proxy of the solar F10.7 cm radio line [19]. The ratio of coronal to photospheric composition ($A_c/A_p$) increases from around 2.3 (2010) to close to 4 (2014). Irradiance measurements allowed to compute the daily averaged ratio of coronal to photospheric composition, or FIP bias ($A_c/A_p$), for the period between April 2010 and May 2014. Notably, $A_c/A_p$ increases when solar activity picks up.

The observed solar composition problem becomes more puzzling when taking into consideration **Figure 1**a, which shows two peaking planetary relationships. Following conventional reasoning, these relationships should not be extant. The day of the measured elemental abundances (given by FIP/BIN in **Figure 1**a) is projected on the corresponding Earth's heliocentric orbital position, and the other to that of Mercury being constrained by Venus' heliocentric orbital position between 20° and 140°. The observed peaking spectral shapes exclude on their own any long-range force [4,5] such as the debated gravitational tidal forces, even though these are extremely feeble [11] and could not cause any visible impact similar to the present one.

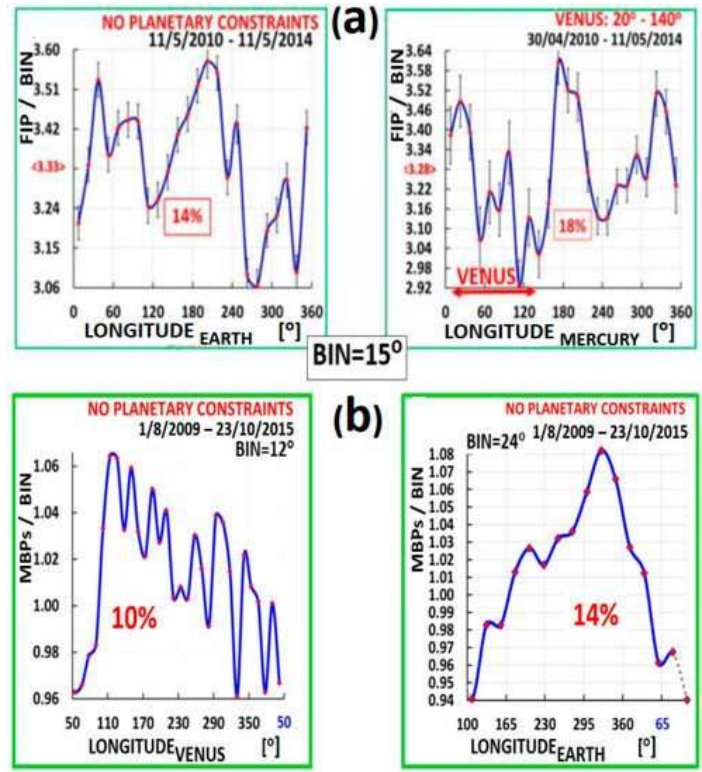

**Figure 1.** (**a**) The dependence of the coronal elemental composition [19] to the photospheric one ($A_c/A_p$), given here by FIP / BIN, is projected on heliocentric longitudes for Earth (*left*) and the combined Mercury—Venus dependence (*right*). The relative min ⇔ max amplitude is 14% and 18%, respectively. (**b**) the dependence of the relative number of magnetic bright points (MBPs) [20] projected on heliocentric longitudes of Venus (*left*) and Earth (*right*). The estimated error per BIN is about 3.5%. This also follows from the rather smooth shape of Earth's spectral shape (*right*). The relative min ⇔ max amplitude is 10% and 14%. See also **Supplementary Materials** with more relevant information.

(b) the solar magnetic bright points (MBPs) [20]. The Sun shows a global magnetic field changing with the 11-year solar cycle. In addition, our host star also harbours small-scale magnetic fields that are often seen as strong concentrations of magnetic flux reaching kG field strengths. **Figure 1**b shows two planetary relationships using a decade-long daily relative number of solar MBPs at the solar disc center being projected in the frame of reference of Venus and Earth.

## 4. Discussion

The detection of the constituents of dark matter remains as one of the central challenges in modern physics. The strongest evidence of DM comes from large scale cosmic observations, while direct searches have failed thus far to provide convincing evidence of it. The large scale observations suggest that the ordinary DM halo in the galaxy is relatively isotropic, at least for the size of the solar system; in the literature, both the co-existence of dark streams (see also **Figure 2**) and the galactic dark disk hypothesis have also been considered [14,21]. The existence of DM streams could explain the puzzling behavior of the active Sun, where there is not yet a clear picture of its workings, e.g., phenomena such as the solar flares and the unnaturally hot corona (see e.g., [13] and ref [6] therein). In this work, we occasionally refer to generic dark candidate constituents as "invisible massive matter", in order to distinguish them from ordinary dark matter such as the celebrated axions and WIMPs.

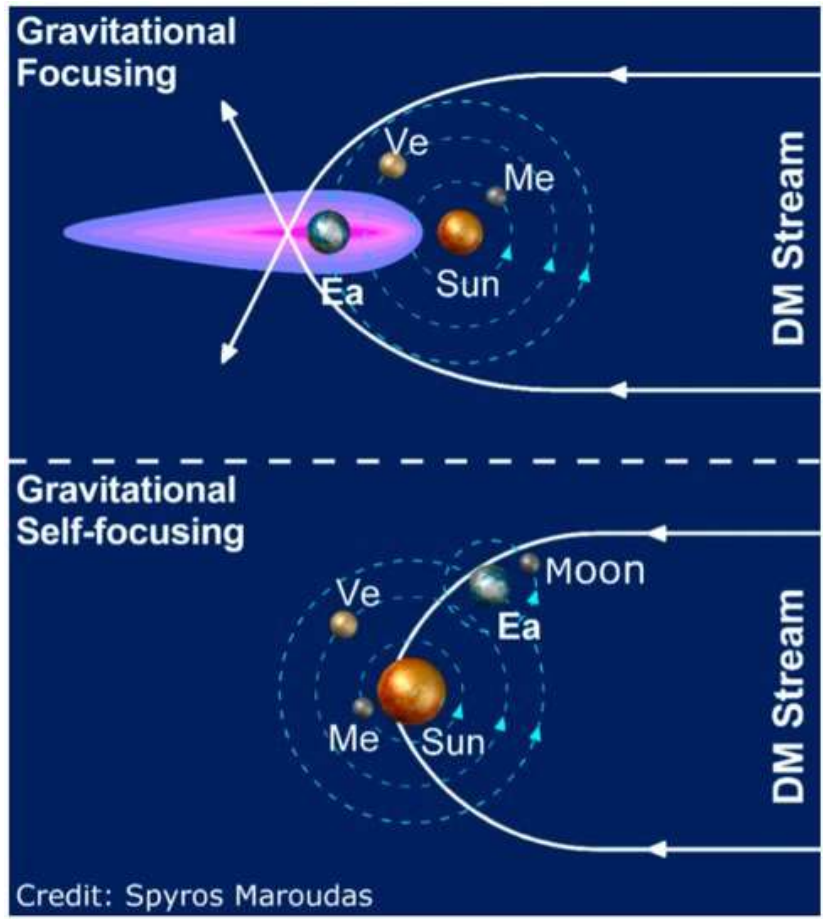

**Figure 2.** Schematic representation of gravitational (self-)focusing effects of DM streams by the Sun, Earth, Venus, Mercury and/or Moon: (*Top*) gravitational focusing effect by the solar system. In this configuration, the galactic center is on the right side and in the opposite direction of the incident DM stream; (Bottom) the self-focusing effect of incident low speed streams reflects the dominating free fall towards the Sun; the flux enhancement increases with ($v_{incident}/v_{escape})^2$. The flux towards Earth can also be gravitationally modulated by the intervening Moon [10]. See also **Supplementary Materials** for additional information related to these scetches.

Of note, the intriguing and as yet unanswered question is whether the motor of the active Sun is entirely of an internal nature, or if it is triggered by some external influence. Here we follow the latter scenario, by assuming that the triggering mechanism is at least partly due to the planetary focusing of some invisible massive matter stream(s) possibly also with a large interaction cross section with ordinary matter.

Furthermore, even if only a portion of the observed solar activity arises from the focusing of some DM streams by the outer planets, straightforward tests of this scenario should be possible. First, the direction of the inferred DM stream in the reference frame of our galaxy could be compared with the stellar halo streams that are now being identified in the Gaia astrometry mission data. These are believed to arise from one or more ancient galaxy mergers with our own Milky Way, and which should carry with them the dark matter halos predicted by current cosmological models [22]. Second, the direction of such a hypothesized DM stream would presumably be fixed, in galactic coordinates, at over at least several tens of parsecs in our solar neighborhood. The study of the time-dependency of the stellar activity of other exoplanet systems in our solar neighborhood would then be expected to correlate with similar planetary focusing also occurring in those systems [23].

## 5. Conclusions

In this work, we have presented accumulating evidence of small-scale observations that show a planetary relationship. The remote gravitational planetary effect impact is extremely feeble and is excluded as the origin behind a plethora of diverse observations. A common viable scenario for such observations is that of the streaming invisible matter, which undergoes planetary gravitational focusing towards the Sun or Earth, enormously enhancing the local flux of DM. This, combined with possible constituents from the dark sector with a large interaction cross-section with ordinary material, can further amplify the impact of focused DM streams.

The suggested scenario follows from the present and previous work. Therefore, it seems reasonable to conclude that DM is occasionally visible, but it has been overlooked to this point, mainly due to the failed dedicated DM searches. Therefore, in future, the strategy of direct DM searches should be changed. Of note, our favorite DM candidates are such as the anti-quark nuggets, magnetic monopoles and dark photons from the hidden sector. Future direct experimental searches should turn their sensors towards such DM constituents.

Similarly, the focusing of DM streams could also occur in other nearby exoplanetary systems, where streaming DM is experienced in the same way as our solar system. Planetary focusing in those systems could be initially investigated by searching for the associated stellar activity as a function of the exoplanetary orbital phases.

**Supplementary Materials**

Supplementary materials are available online at
**https://www.mdpi.com/article/10.3390/ECU2021-09313/s1**.

**Author Contributions**

All authors have contributed in preparing this work, writing and editing the manuscript, and have also contributed in previous work upon which builds the present paper. All authors have read and agreed to the published version of the manuscript.

**Funding**

For M. M. this research is co-financed by Greece and the European Union (European Social Fund—ESF) through the Operational Programme "Human Resources Development, Education and Lifelong Learning" in the context of the project "Strengthening Human Resources Research Potential via Doctorate Research" (MIS-5000432), implemented by the State Scholarships Foundation (IKY).

**Data Availability Statement,**

David Brooks kindly provided the elemental composition data (Fig. 1a). The author D.U. should be contacted about the data used for the analysis (Fig. 1b).

**Acknowledgments,**   For Y. K. Semertzidis this work was supported by IBS-R017-D1. We thank David Brooks for kindly providing the elemental abundances data.